\documentclass[manuscript, noacm]{acmart}

\renewcommand\footnotetextcopyrightpermission[1]{}

\AtBeginDocument{%
  }

\acmConference[CONSEQUENCES @ RecSys]{}{2022}{Seattle, WA, USA}

\usepackage[ruled,vlined,linesnumbered]{algorithm2e}
\SetKwComment{Comment}{$\triangleright$\ }{}

\def\eqref#1{Equation~\ref{#1}}			% reference to equation
\def\figref#1{Figure~\ref{#1}}			% reference to figure
\def\algref#1{Algorithm~\ref{#1}}		% reference to algorithm
\def\dffref#1{Definition~\ref{#1}}		% reference to definition
\newtheorem{dff}{Definition}

\newcommand{\pag}{\mathcal{G}}

\newcommand{\ipath}[2]{\Pi(#1, #2)}

\DeclareMathOperator{\sindep}{Ind}  % statistical conditional independence (or oracle)

\makeatletter
\newcommand*{\indep}{%
  \mathbin{%
    \mathpalette{\@indep}{}%
  }%
}
\newcommand*{\nindep}{%
  \mathbin{%                   % The final symbol is a binary math operator
    \mathpalette{\@indep}{\not}% \mathpalette helps for the adaptation
  }%
}
\newcommand*{\@indep}[2]{%
  \sbox0{$#1\perp\m@th$}%        box 0 contains \perp symbol
  \sbox2{$#1=$}%                 box 2 for the height of =
  \sbox4{$#1\vcenter{}$}%        box 4 for the height of the math axis
  \rlap{\copy0}%                 first \perp
  \dimen@=\dimexpr\ht2-\ht4-.2pt\relax
  \kern\dimen@
  {#2}%
  \kern\dimen@
  \copy0 %                       second \perp
} 
\makeatother

\newcommand{\recsys}{\mathcal{M}}
\newcommand{\A}{\boldsymbol{A}}
\newcommand{\sess}{\boldsymbol{S}}
\newcommand{\replacement}{\boldsymbol{R}}
\newcommand{\recommendation}{\Tilde{I}}
\newcommand{\rec}{\recommendation_{n+1}}
\newcommand{\reppot}{\recommendation_{n+1}'}

\newcommand\blfootnote[1]{%
  \begingroup
  \renewcommand\thefootnote{}\footnote{#1}%
  \addtocounter{footnote}{-1}%
  \endgroup
}

\begin{document}

\title{CLEAR: Causal Explanations from Attention in Neural Recommenders}

\author{Shami Nisimov}
\affiliation{%
  \institution{Intel Labs}
}
\email{shami.nisimov@intel.com}

\author{Raanan Y. Rohekar}
\affiliation{%
  \institution{Intel Labs}
}
\email{raanan.yehezkel@intel.com}

\author{Yaniv Gurwicz}
\affiliation{%
  \institution{Intel Labs}
}
\email{yaniv.gurwicz@intel.com}

\author{Guy Koren}
\affiliation{%
  \institution{Intel Labs}
}
\email{guy.koren@intel.com}

\author{Gal Novik}
\affiliation{%
  \institution{Intel Labs}
}
\email{gal.novik@intel.com}

\begin{abstract}
  We present CLEAR, a method for learning session-specific causal graphs, in the possible presence of latent confounders, from attention in pre-trained attention-based recommenders. These causal graphs describe user behavior, within the context captured by attention, and can provide a counterfactual explanation for a recommendation. In essence, these causal graphs allow answering ``\emph{why}'' questions uniquely for any specific session. Using empirical evaluations we show that, compared to naively using attention weights to explain input-output relations, counterfactual explanations found by CLEAR are shorter and an alternative recommendation is ranked higher in the original top-k recommendations.
\end{abstract}

\maketitle

\section{Introduction}

An automated system that provides recommendations to humans should be explainable\blfootnote{Causality, Counterfactuals and Sequential Decision-Making for Recommender Systems (CONSEQUENCES) workshop at RecSys 2022, Seattle, WA, USA.}. An explanation to why a specific recommendation was given to a human in a tangible way can lead to greater trust in the recommendation, and greater human engagement with the automated system. In recent years, recommenders based on deep neural networks have demonstrated state-of-the-art accuracy \cite{zhang2019deep, he2017neural, xin2019relational}. Among these, attention-based recommenders \cite{kang2018self, sun2019bert4rec} are based on an architecture (Transformer) \cite{vaswani2017attention} that scale well with model and data size. However, it is not clear how to interpret these models and how to extract explanations meaningful for humans. A common approach is to use the attention matrix, computed within these models, to learn about input-output relations for providing explanations \cite{seo2017interpretable, chen2019personalized, chen2018neural}. These often rely on the assumption that inputs having high attention values influence the output \cite{xu2015show, choi2016retain}. In this paper, we claim that this assumption considers only marginal statistical dependence and ignores conditional independence relations. Moreover, we claim that a post-hoc view is more suitable for explaining a recommendation.

It was previously claimed that attention cannot be used for explanation \cite{jain2019attention}. And recently it was claimed that attention cannot be used for providing counterfactual explanations for recommendations made by neural recommenders \cite{tran2021counterfactual}. In a paper contradicting \citet{jain2019attention}, it was shown that explainability is task dependent \cite{wiegreffe2019attention}. \citet{lipton2019mythos} defines human understanding of a model, and post-hoc explainability as two distinct notions. 

\begin{figure}[th]
  \centering
  \includegraphics[width=\linewidth]{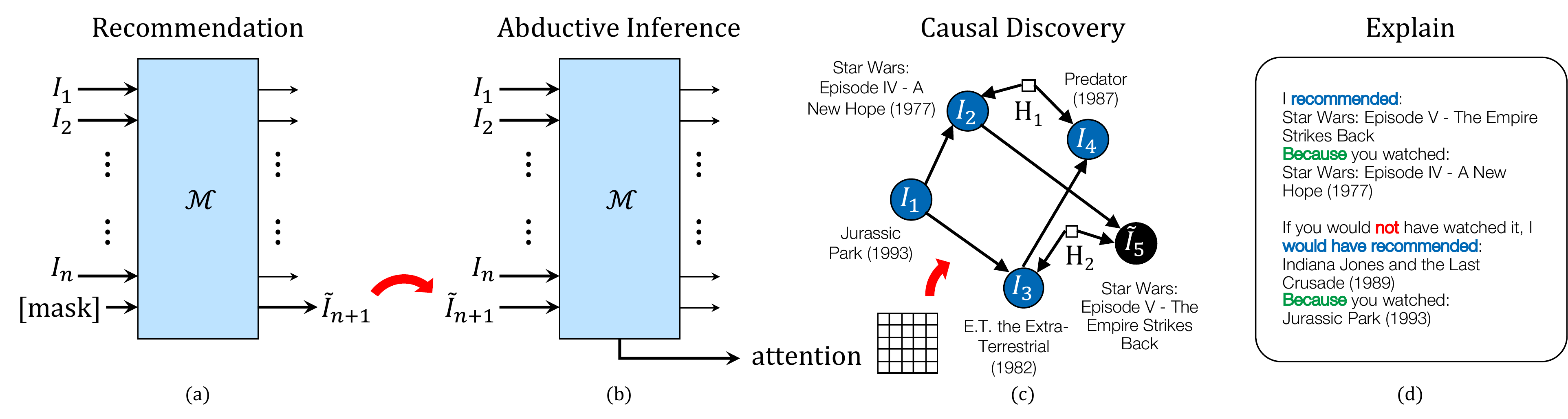}
  \caption{An overview of the presented approach. As an example, $\mathcal{M}$ is a pre-trained BERT4Rec recommender \cite{sun2019bert4rec}. (a) Given a set of $n$ items, $\{I_1,\ldots,I_n\}$ that the user interacted with, and the $n+1$ item masked, an attention based recommender $\mathcal{M}$ recommends $\Tilde{I}_{n+1}$. Next, in the abductive inference stage (b) the recommendation is treated as input in addition to the observed user-item interactions, that is, input: $\{I_1,\ldots,I_n, \Tilde{I}_{n+1}\}$, and the attention matrix is extracted. (c) A causal structure learning stage in which the attention matrix is employed for testing conditional independence in a constraint-based structure learning algorithm. The result is a Markov equivalence class of causal graphs with possibly hidden confounders. Here, as an example, a causal DAG is given, where $H_1$ and $H_2$ are hidden confounders. (d) A counterfactual explanation reasoned from the causal graph.} 
  \Description{The three main stages of the algorithm drawn in three columns. In the first and second stages a neural recommender is drawn as rectangle with inputs and outputs. In the third stage an example causal graph is drawn.}
  \label{fig:approach}
\end{figure}

In this paper we learn a causal graph for each session, and suggest it as a mean for human understanding of the model. We do not claim that attention-based recommenders learn causal graphs for sessions. Instead, we consider the learned causal graph as a projection of the model. Next, by considering the recommendation as part of an imaginary session, we extract an explanation set from the causal graph. This set is validated by omitting it from the original session and feeding the edited session into the recommender, resulting in an alternative recommendation (which can be explained in a similar manner). 
An overview of the presented approach is given in \figref{fig:approach}. The pseudo-code for identifying an explanation for any specific session is given in \algref{alg:Alg} and detailed in the next sections.

\begin{algorithm}%[H]
\SetKwInput{KwInput}{Input}                % Set the Input
\SetKwInput{KwOutput}{Output}              % set the Output
\SetKw{Break}{break}
\DontPrintSemicolon
  
  \BlankLine
  
  \KwInput{
    \\\quad $\sess$: a session consisting of items that user interacted with $\sess=\{I_1,\ldots,I_n\}$
    \\\quad $\recsys$: a pre-trained attention-based neural recommender
    \\\quad $\rec$: recommended item (by $\recsys$)
    \\\quad $\alpha$: significance level for CI-testing
    \\\quad $\replacement$: (optional) a pool of possible replacement recommendations (e.g., items in the top-$k$ recommendation slate)
    }
    
    \BlankLine
    
  \KwOutput{
  \\\quad $\boldsymbol{E}$: an explanation set for the recommendation
  \\\quad $\reppot$: an alternative recommendation (had the explanation not in the session)
  \\\quad $\pag$: a causal graph of $\sess + \rec$}

% Set Function Names
  \SetKwFunction{FMain}{Main}
  \SetKwFunction{FindExp}{FindExplanation}
  \SetKwFunction{ICD}{ICD}
  \SetKwFunction{FIter}{CDIteration}
  \SetKwFunction{FPDSepRange}{PDSep\_r}

\BlankLine\BlankLine\BlankLine

% Write Function with word ``Function''
  \SetKwProg{Fn}{Function}{:}{}
  \Fn{\FMain{$\sess$, $\recsys$, $\rec$, $\replacement$}}{
        $\Tilde{\sess} \leftarrow \sess + \rec$ \Comment*{concatenate recommendation with observed session}
        forward pass: $\recsys(\Tilde{\sess})$\;
        get $\Tilde{\A}$, the attention matrix in the last layer of $\recsys(\Tilde{\sess})$\;
        % $\Tilde{\A} \leftarrow$ \GetAtten($\recsys(\Tilde{\sess})$) \Comment*{get values from the last attention layer }
        % $\boldsymbol{K} \leftarrow \Tilde{\A} \Tilde{\A}^\mathrm{T}$ \;
        define correlation matrix $\boldsymbol{\rho}_{\Tilde{\sess}} : \boldsymbol{\rho}_{\Tilde{\sess}}({i,j})=\boldsymbol{K}({i,j}) / \sqrt{\boldsymbol{K}({i,i})\cdot \boldsymbol{K}({j,j})}$, where $\boldsymbol{K} = \Tilde{\A} \Tilde{\A}^\mathrm{T}$ \;
        define $\sindep(\boldsymbol{\rho}_{\Tilde{\sess}})$ : a conditional independence test based on partial correlation \;
        $\pag \leftarrow$ \ICD($\sindep(\boldsymbol{\rho}_{\Tilde{\sess}})$, $\alpha$) \Comment*{session-specific causal-discovery}
        $\boldsymbol{E}, \reppot \leftarrow$ \FindExp($\pag$, $\rec$, $\recsys$, $\sess$, $\replacement$) \Comment*{find an explanation set for the recommendation}
        \KwRet $\boldsymbol{E}$, $\reppot$, $\pag$\;
  }

  \BlankLine\BlankLine\BlankLine

  \SetKwProg{Fn}{Function}{:}{\KwRet}
  \Fn{\FindExp{$\pag$, $\rec$, $\recsys$, $\sess$, $\replacement$}}{
        create $\mathcal{T}$: a PI-tree for $\rec$ given $\pag$ \;
        \For{$r$ ~in~ $\{1, \ldots, n-1\}$}{
            create $\mathcal{E}$, subsets of $\mathbf{Nodes}(\mathcal{T})\smallsetminus \rec$ such that $\forall \boldsymbol{E}\in\mathcal{E}$, $\boldsymbol{E}$ is a PI-set on $\rec$ in $\pag$ with respect to $r$. \;
            \For{$\boldsymbol{E}\in\mathcal{E}$}{
                $\sess' \leftarrow \sess \smallsetminus \boldsymbol{E}$\;
                $\reppot \leftarrow \recsys(\sess')$ \Comment*{get an alternative recommendation}
                \If{$\reppot \neq \rec$ and $\reppot \in \replacement$}{
                    \KwRet $\boldsymbol{E}$, $\reppot$ \Comment*{return immediately ensuring smallest explanation set}
                }
            }
        }
        \KwRet $\emptyset$, $\emptyset$ \Comment*{no explanation found at the current significance level (consider a higher $\alpha$)}
    }

  \caption{CLEAR: CausaL Explanations from Attention in neural Recommenders}
  \label{alg:Alg}
\end{algorithm}

\section{Causal Structure Learning from Attention}

We assume that the human decision process, for selecting which items to interact with, consists of multiple decision pathways that may diverge and merge over time. Moreover, they may be influenced by latent confounders along this process. Formally, we assume that the decision process can be modeled by a causal DAG consisting of observed and latent variables. Here, the observed variables are user-item interactions $\{I_1,\ldots,I_n\}$ in a session $S$, and latent variables $\{H_1, H_2, \ldots\}$ represent unmeasured influences on the user's decision to interact with a specific item. Examples for such unmeasured influences are user intent and previous recommendation slates presented to the user.

\subsection{Causal Structure Learning}

Causal structure learning (causal discovery) from observed data alone requires placing certain assumptions. Here we assume the causal Markov \cite{pearl2009causality} and faithfulness \cite{spirtes2000} assumptions, and do not place parametric assumptions on the distribution of the data. Under these assumptions, constraint-based methods use tests of conditional independence (CI-tests) to learn the causal structure \cite{pearl1991theory, spirtes2000, colombo2012learning, claassen2013learning, rohekar2021iterative, rohekar2018brai, nisimov2021improving}. The resulting graphs represent an equivalence class in the form of a partial ancestral graph (PAG) \cite{richardson2002ancestral, zhang2008completeness}. A PAG represents a set of causal graphs that cannot be refuted given the data. There are three types of edge-marks (at some node $X$): an arrow-head '{---}>$X$', a tail '{---}--$X$', and circle `---o $X$' which represent an edge-mark that cannot be determined given the data. Throughout the paper we refer to PAG as a causal graph such that reasoning from it is consistent with every member in the equivalence class it represents. In this paper, we use the ICD algorithm \cite{rohekar2021iterative} for learning a PAG, as it is sound and complete, and was demonstrated to be more efficient and accurate than other algorithms. Nevertheless, the presented approach is not restricted to it.

\subsection{Attention for Conditional Independence Testing}

Every constraint-based causal discovery algorithm uses a CI-test for deciding if two nodes are independent given a conditioning set of nodes. Commonly partial correlation is used for CI-testing between continuous, normally distributed variables with linear relations. This test requires only a pair-wise correlation matrix (marginal dependence) for evaluating partial correlations (conditional dependence). 

We use partial-correlation CI testing after evaluating the correlation matrix from the attention matrix. Specifically, we assume a mapping of the user-item interactions to an RKHS. We assume the attention matrix $\boldsymbol{A}$, evaluated in the last attention layer (\algref{alg:Alg}-lines 2--4), to represents functions in this RKHS. We define covariance $K=A\cdot A^{\mathrm{T}}$ and evaluate correlation coefficients $\rho_{i,j}=K_{i,j} / \sqrt{K_{i,i}\cdot K_{j,j}}$ (\algref{alg:Alg}-line 5). 

Unlike kernel-based CI tests \cite{bach2002kernel, fukumizu2004dimensionality, gretton2005measuring, gretton2005kernel, sun2007kernel, zhang2011kernel}, we do not need to explicitly define the kernel nor do we need to compute the kernel matrix, as it is readily available by a single forward-pass in the Transformer (\algref{alg:Alg}-line 3). This implies the following. Firstly, our CI-testing function is inherently learned during the training stage of a Transformer, by that enjoying the efficiency in learning complex models from large datasets. Secondly, since attention is computed for each input uniquely, CI-testing is unique to that specific input. Using this CI-testing function we call a causal discovery algorithm and obtain a causal graph (\algref{alg:Alg}-lines 6,7).

\section{Finding Possible Explanations from a Causal Graph}

Given a causal graph, various ``\emph{why}'' questions can be answered \cite{pearl2018book}. In this paper we follow \cite{tran2021counterfactual} explaining a recommendation using the user's own actions (user-item interactions). That is, provide the minimal set of user-item interactions that led to a specific recommendation and provide an alternative recommendation. To this end, we consider a causal graph that includes the recommendation as part of an imaginary session, and define the following.  

\begin{dff}[PI-path]\label{dff:i-path}
    A potential influence path from $A$ to $Z$ in PAG $\pag$, is a path $\ipath{A}{Z}=\langle A,\ldots,Z\rangle$, such that for every sub-path $\langle U,V,W \rangle$ of $\ipath{A}{Z}$, where there are arrow heads into $V$ and there is no edge between $U$ and $W$ in $\pag$.
\end{dff}

Essentially, a PI-path ensures dependence between its two end points when conditioned on every node on the path (a single edge is a PI-path).

\begin{dff}[PI-tree]\label{dff:i-tree}
    A potential influence tree for $A$ given causal PAG $\pag$, is a tree $\mathcal{T}$ rooted at $A$, such that there is a path from $A$ to $Z$ in $\mathcal{T}$, $\langle A, V_1,\ldots, V_k, Z\rangle$, if and only if there is an PI-path $\langle A,V_1,\ldots, V_k,Z \rangle$ in $\pag$.
\end{dff}

\begin{dff}[PI-set]\label{dff:ip-conditions}
    A set $\boldsymbol{E}$ is a potentially influencing set (PI-set) on item $A$, with respect to $r$, if and only if:
    \begin{enumerate}
        \item $|\boldsymbol{E}| = r$
        \item $\forall E\in\boldsymbol{E}$ there exists a PI-path, $\ipath{A}{E}$ such that $\forall V\in\ipath{A}{E}, V\in\boldsymbol{E}$
        \item $\forall E\in\boldsymbol{E}$, $E$ temporally precedes $A$.
    \end{enumerate}
\end{dff}

\begin{figure}[h]
  \centering
  \includegraphics[width=0.85\linewidth]{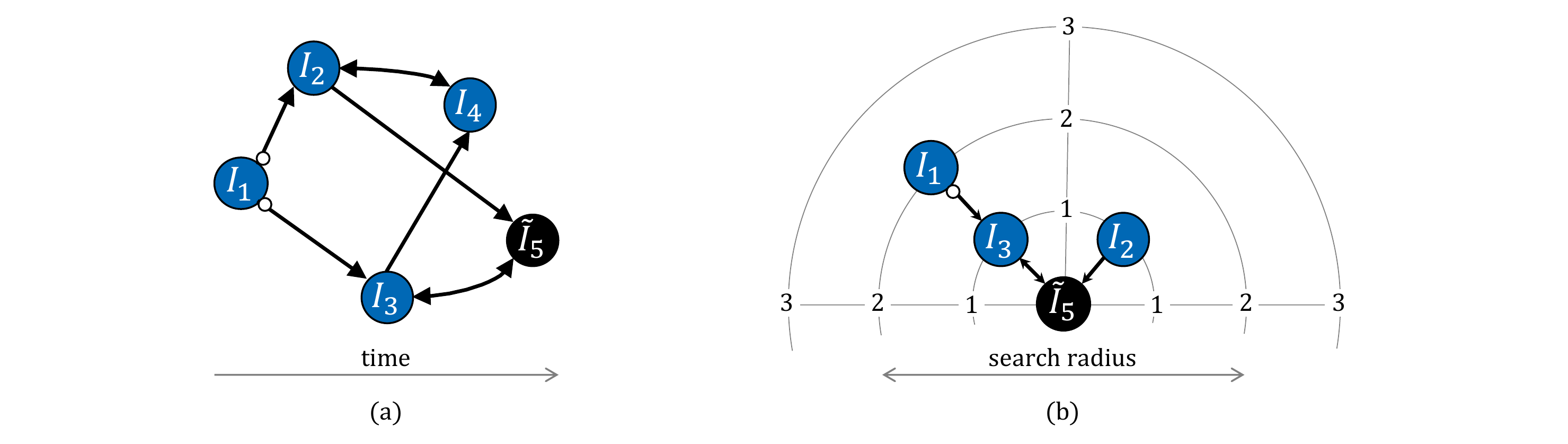}
  \caption{An example for (a) a causal graph (PAG), and (b) the nodes that influence the recommendation $\Tilde{I}_5$ having a PI-tree structure, depicted on coordinates where the radius axis is the distance from the recommendation. In \algref{alg:Alg}-line 12, the algorithm iterates over radius values, starting with 1. In the first iteration, the sets $\{I_2\}$ and $\{I_3\}$ are tested if any of them qualifies as an explanation. If both sets fail, the search radius increases to 2 and the sets $\{I_2, I_3\}$ and $\{I_1, I_3\}$ are tested (set $\{I_2, I_3\}$ is tested first as it members have smaller average distance from $\Tilde{I_5}$). If both fail, $r=3$ and the last tested set is $\{I_1, I_2, I_3\}$.}
  \Description{A causal graph with hidden variables, and a diagram with pseudo polar coordinates describing nodes that affect the target node.}
  \label{fig:search-radius}
\end{figure}

An example for identifying PI-sets is given in \figref{fig:LatentMarkovBlanket}. Although a PI-set identify nodes that are conditionally dependent on the recommendation, it may not be a minimal explanation for the specific recommendation. That is, omitting certain items from the session alters the graph and may render other items independent of the recommendation. Hence, we provide an iterative procedure (``FindExplanation'' in \algref{alg:Alg}-lines 10--18), motivated by the ICD algorithm \cite{rohekar2021iterative}, to create multiple candidate explanations by gradually increasing the value of $r$. See an example in \figref{fig:search-radius}. Note that from conditions (1) and (2) of \dffref{dff:ip-conditions}, $r$ effectively represents the maximal search radius from the recommendation (in the extreme case, the items of the PI-set lie on one PI-path starting at the recommendation). The search terminates as soon as a set qualifies as an explanation (\algref{alg:Alg}-line 17). That is, as soon as the recommender provides a different recommendation. If no explanation is found (\algref{alg:Alg}-line 19) a higher value of $\alpha$ should be considered for the algorithm.

\section{Empirical Evaluation}
 
For empirical evaluation, we use the BERT4Rec recommender \cite{sun2019bert4rec}, pre-trained on the MovieLens 1M dataset \cite{movielens} and estimate several measures to evaluate the quality of reasoned explanations. We compare CLEAR to a baseline (Atten.) attention-based algorithm \cite{tran2021counterfactual}, that uses the attention weights directly in a hill-climbing search to suggest an explaining set. On average, the number of BERT4Rec forward-passes executed per-session by Atten. was 3 time greater than by CLEAR.

\subsection{Influence of the explaining set on replacement recommendation}

\begin{figure}[h]
  \centering
  (a) \includegraphics[width=0.4\linewidth]{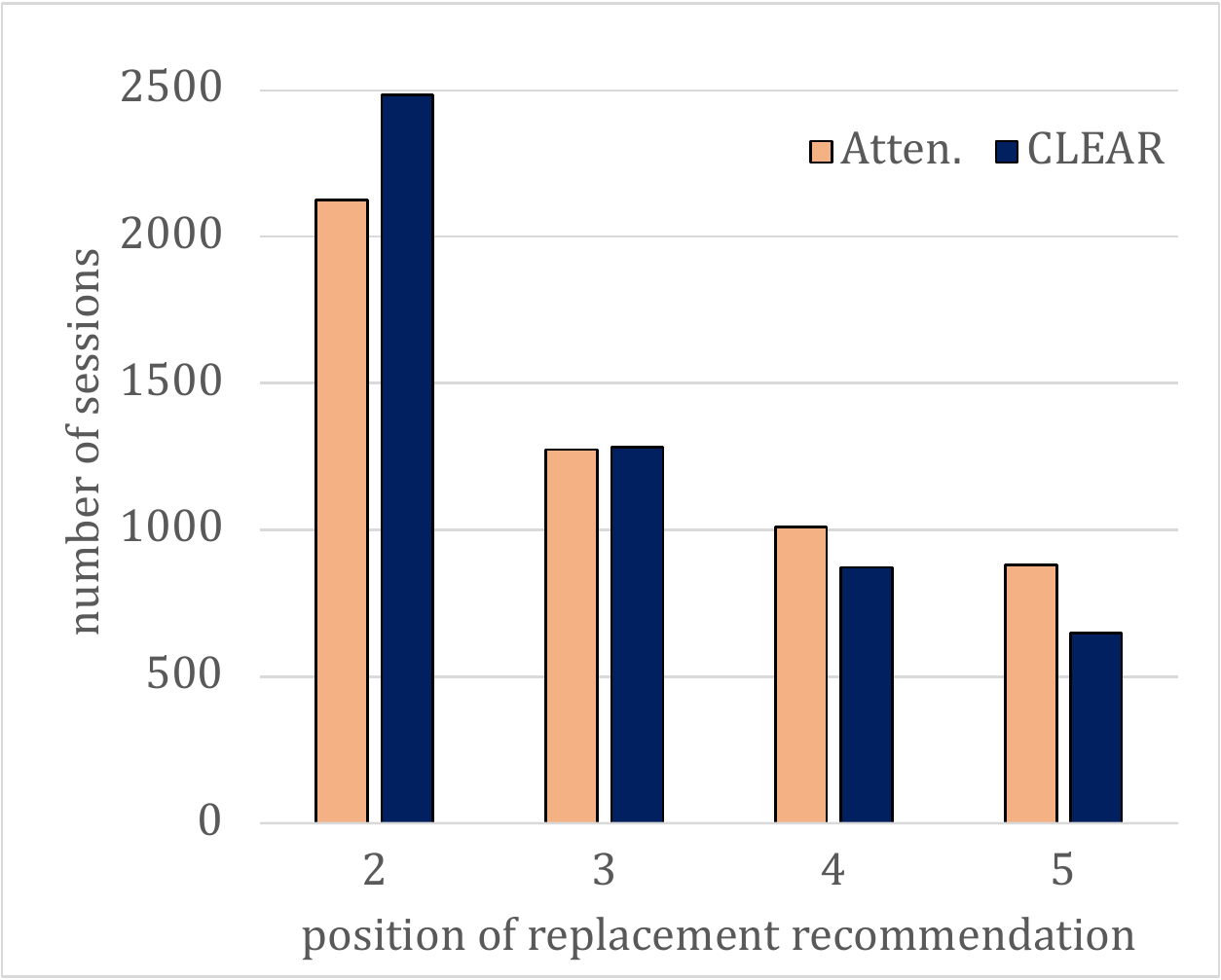}\qquad\qquad
  (b) \includegraphics[width=0.4\linewidth]{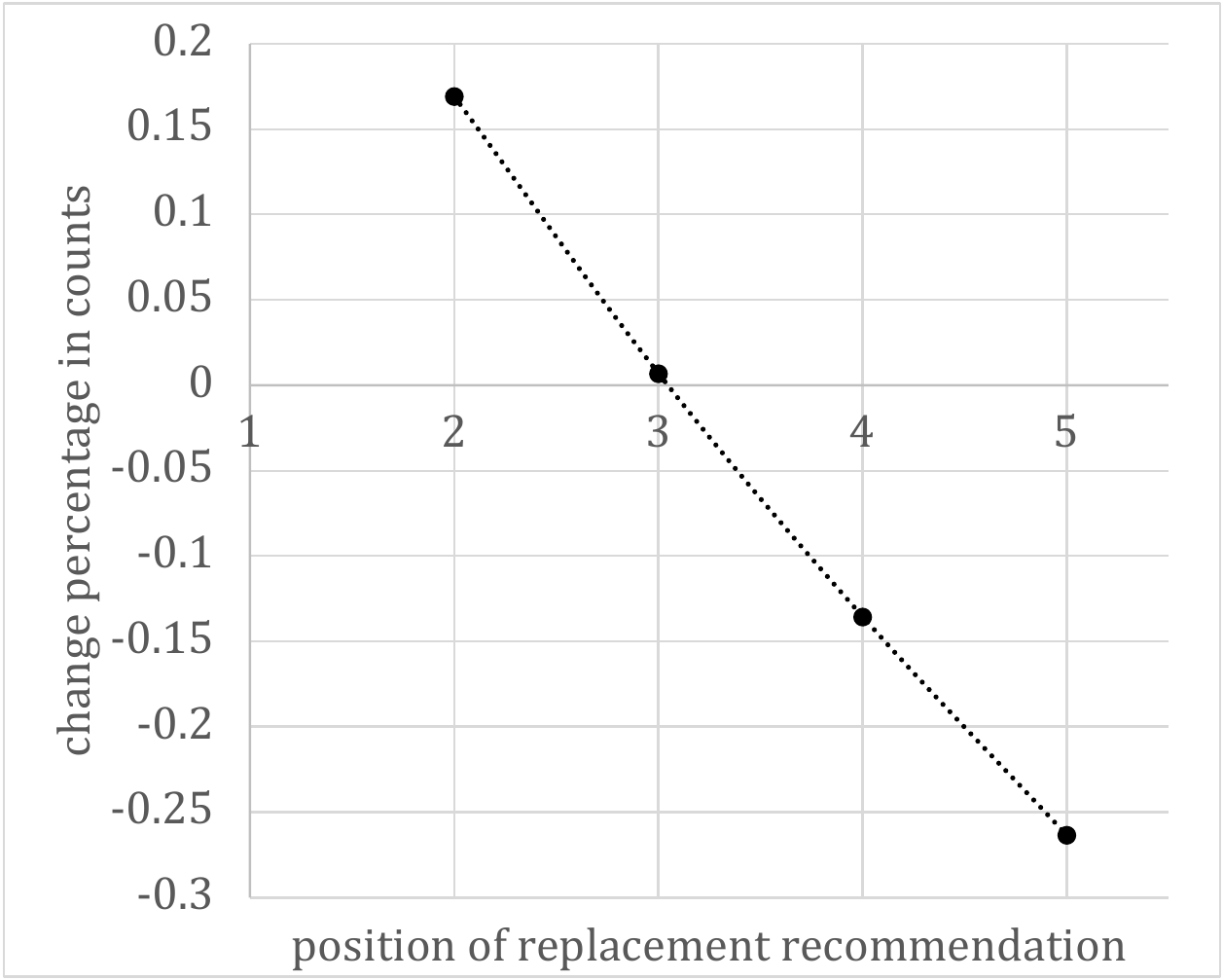}
  \caption{Positions of the replacement recommendations within the original top-5 recommendations. (a) Position distribution of the replacements: compared to Attention, CLEAR recommends replacements which are ranked higher in the original top-5 recommendations; (b) The relative gain in the number of sessions for each position, achieved by CLEAR. There is a trend line indicating higher gains for CLEAR at lower positions (replacements are more aligned with the original top-5 recommendation).}
  \label{fig:positions}
\end{figure}

Given a session consisting of items that a user interacted with, $\sess=\{I_1,\ldots,I_n\}$, a neural recommender suggests $\rec$, the 1\textsuperscript{st} item in the top-5 recommendations list. CLEAR finds the smallest explaining set within $\sess$ that influenced the recommendation of $\rec$. As a consequence, discarding this explaining set from that session should prevent that $\rec$ from being recommended again, and instead a new item should be recommended in replacement. Optimally, the explaining set should influence the rank of only that 1\textsuperscript{st} item (should be downgraded), but not the ranks of the other recommendations in the top-5 list. This requirement implies that the explaining set is unique and important for the isolation and counterfactual explanation of only that 1\textsuperscript{st} item, whereas the other items in the original top-5 list remain unaffected, for the most part. It is therefore desirable that after discarding the explaining set from the session, the new replacement item would be one of the original (i.e. before discarding the explaining set) top-5 recommendation. To quantify this, CLEAR finds the replacement recommended item for each session, and \figref{fig:positions}(a) shows the distribution of their positions within their \emph{original} top-5 recommendations list. It is evident that compared to the baseline Attention method, CLEAR recommends replacements that are ranked higher (lower position) in the original top-5 recommendations list. In a different view, \figref{fig:positions}(b) shows the relative gain in the number of sessions for each position, achieved by CLEAR compared to Attention. There is a trend line indicating higher gains for CLEAR at lower positions, i.e. the replacements are more aligned with the original top-5 recommendations. CLEAR is able to isolate a minimal explaining set that influence only the 1\textsuperscript{st} item from the original recommendations list.

\subsection{Minimal explaining set}

\begin{figure}[h]
  \centering
  (a) \includegraphics[width=0.4\linewidth]{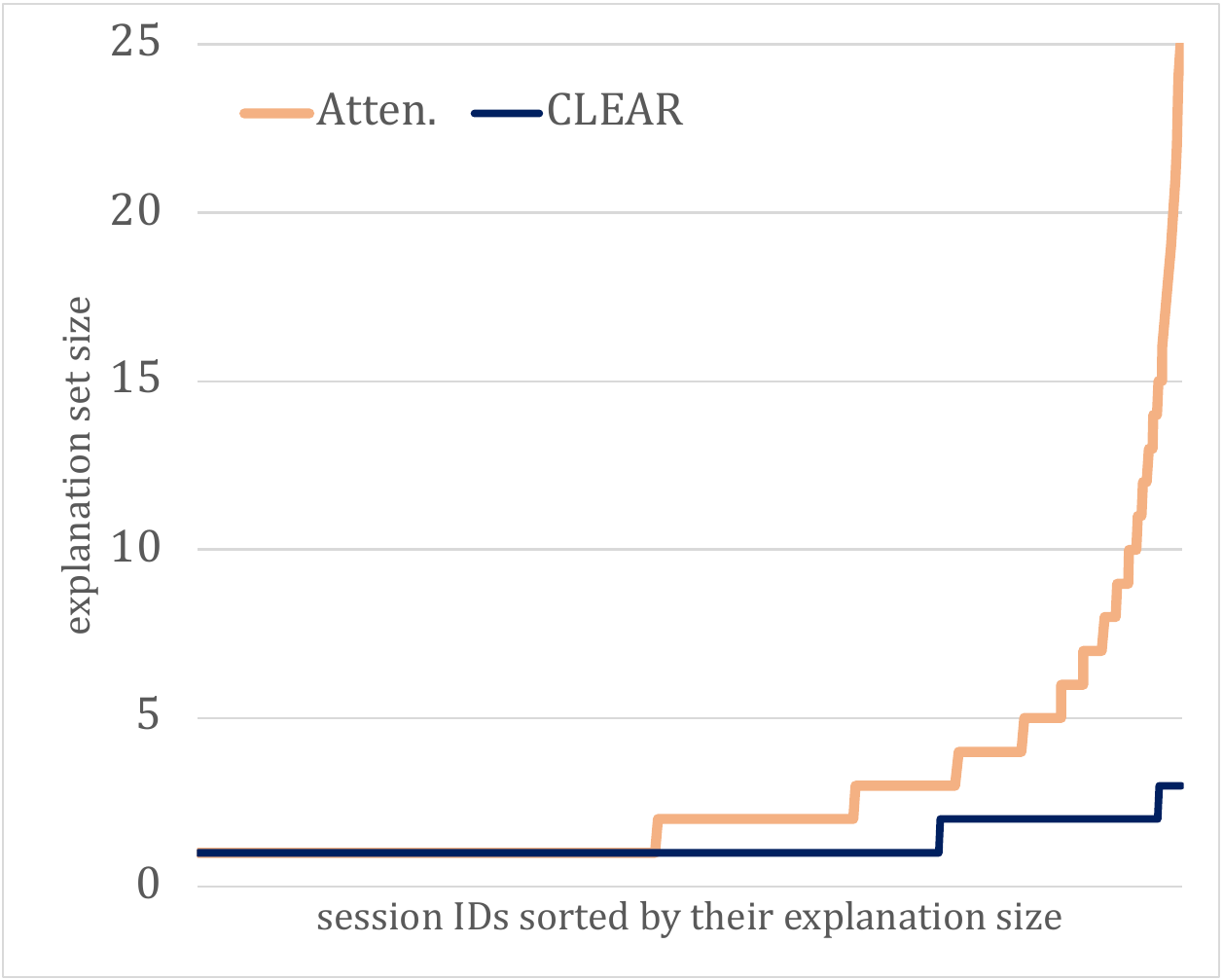}\qquad\qquad
  (b) \includegraphics[width=0.4\linewidth]{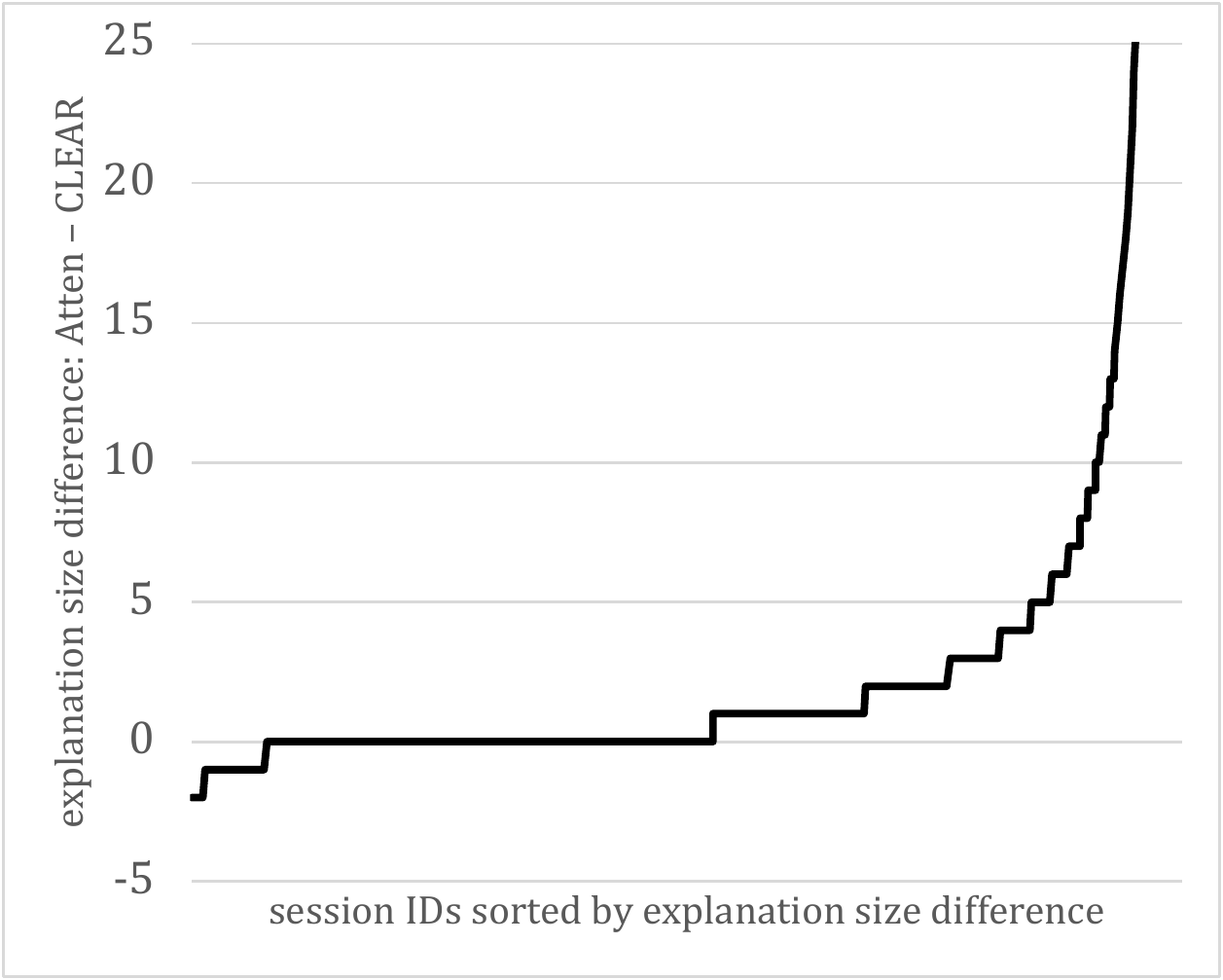}
  \caption{Explaining set size. (a) Comparison between CLEAR and Attention of the explaining set size (sorted by size) for the various sessions. Set sizes found by CLEAR are smaller. (b) The difference between set sizes found by Attention and CLEAR, presented for each individual session. Positive values are in favor of CLEAR.}
  \label{fig:set_size}
\end{figure}

An important requirement is having the explaining set minimal in number of items. The reason for that is threefold: (1) an explaining set (for the 1\textsuperscript{st} item) that contains less items in it, potentially contains less causal connections with items outside this set, and therefore removal of this set from the session (in a later stage) is less likely to influence the ranking of items outside the set, e.g. the top-5 items. (2) In addition, it is more complicated and less interpretable for humans to grasp the interactions and interplay in a set that contains many explaining items. (3) In the spirit of occum's razor, when faced with a few possibles explanations, the simpler one is the one most likely to be true. \figref{fig:set_size}(a) compares the explaining set size for the various sessions between CLEAR and Attention. It is evident that the set sizes found by CLEAR are smaller. \figref{fig:set_size}(b) shows the difference between the explaining set sizes found by Attention and CLEAR, presented for each individual session. Approximately half of the sessions are with positive values, indicating smaller set sizes for CLEAR, zero values shows equality between the two, and only 8\% of the sessions are with negative values, indicating smaller set sizes for Attention.

\section{Main Conclusions}

We presented CLEAR, an algorithm for learning a causal graph under a specific session context, and demonstrated its usefulness in finding counterfactual explanations for recommendations. We expect learned causal graphs to be able to answer a myriad of causal queries \cite{pearl2009causality}, among these are personalized queries that can allow a richer set of recommendations. For example, assuming that human decision process consists of multiple pathways, by identifying independent causal paths, recommendations can be provided for each one independently. 

An interesting insight is that the only source of errors in CLEAR is errors in the attention matrix learned by the recommender. Since in recent years it was shown that Transformer models scale well with model and training data sizes, we expect CLEAR to be more accurate. 

\bibliographystyle{ACM-Reference-Format}
\bibliography{clear}

\appendix

\section{Additional Explanatory Figures}

This section includes additional explanatory figures for better clarity.

\begin{figure}[h]
  \centering
  \includegraphics[width=\linewidth]{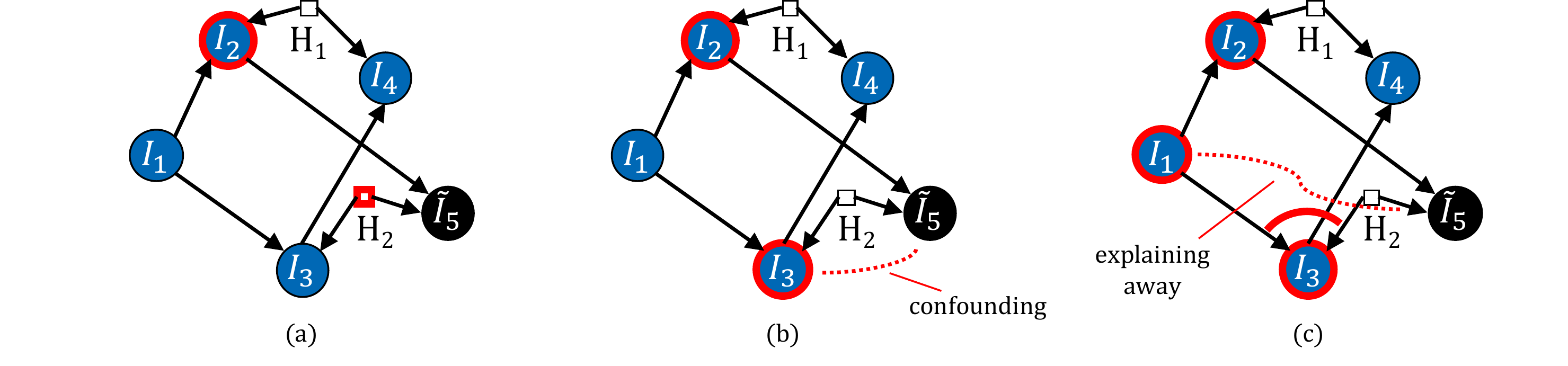}
  \caption{An example for a PI-set on $\Tilde{I}_5$. Latent nodes are represented by squares. Nodes included in the PI-set are circled with a thick red line. Indirect dependency is marked with a dashed red line. (a) The smallest PI-set for a fully observed DAG is the Markov blanket $\{I_2, H_2\}$. However, $H_2$ is latent. (b) Nodes ${I_2, I_3}$ are included in the PI-set as they have direct effect on $\Tilde{I}_5$, where $I_5$ depends on $I_3$ due to the latent confounder $H_2$. However, $I_3$ is a collider, hence including it in the PI-set results in explaining-away and the PI-set is not complete. (c) A complete PI-set $\{I_1, I_2, I_3\}$. Including $I_3$ creates an indirect dependence between $I_5$ and $I_1$ (a PI-path), which requires including $I_1$ as well. Note that the distance from $I_5$ to $I_1$ is two edges in the graph over observed variables.}
  \label{fig:LatentMarkovBlanket}
\end{figure}

\end{document}